\documentclass[11pt]{article}

\usepackage[margin=1in]{geometry}
\usepackage{authblk}
\usepackage{longtable}
\usepackage{array}
\usepackage{calc}
\usepackage{xurl}
\usepackage{soul}
\usepackage{tabularx}
\usepackage{booktabs}
\usepackage{graphicx}
\usepackage{microtype}
\usepackage{indentfirst} 
\usepackage[hidelinks]{hyperref}
\title{Designing Human--AI Collaborative Science: The 9A Framework}

\author[1]{Zihan Zheng}
\author[2]{Linyue Pan}
\author[3]{Xin Lan}
\affil[1]{University of Illinois Urbana--Champaign, Urbana, Illinois, USA}
\affil[2]{Tsinghua University, Shenzhen, Guangdong, China}
\affil[3]{Michigan State University, East Lansing, Michigan, USA}
\date{}

\begin{document}

\maketitle

\begin{abstract}
Artificial intelligence (AI) for Science is often described as a progression from human-directed assistance toward increasingly autonomous AI scientists. Human--AI collaboration offers a further mode in which human and AI contributions substantively shape scientific reasoning, judgments, and the course of inquiry. Existing research addresses work allocation, human--AI coordination, and system evaluation, but how these aspects interact to shape collaborative scientific systems remains insufficiently articulated. We develop the \emph{9A Framework} for Human--AI Collaborative Science (HACS) to connect these considerations across three layers. The \emph{organization} layer specifies what scientific activities are required (Activity), who performs them (Actor), and who holds authority over consequential decisions (Authority). The \emph{coordination} layer addresses how participants access relevant research information (Awareness), how scientific work advances through iterative proposal, execution, examination, revision, and resolution (Accountability), and when collaborative arrangements should be adjusted (Adaptability). The \emph{evaluation} layer establishes what constitutes effective system performance (Achievement), how comparisons can be designed to identify the effects of organizational and coordination choices (Attribution), and what evidence supports the resulting conclusions (Assurance). Together, these layers frame HACS as a system-design problem, providing a basis for designing collaborative scientific workflows and for empirically comparing their effectiveness.
\end{abstract}


\section{Introduction}\label{introduction}

Artificial intelligence is becoming an increasingly capable participant in scientific work. Contemporary systems can retrieve literature, write and debug code, analyze data, generate hypotheses, design experiments, operate tools, and help draft scientific reports \cite{ref07}. More integrated systems now connect several of these activities across substantial portions of a research workflow \cite{ref01,ref02,ref05,ref06}. As AI takes on a larger role in scientific inquiry, how joint human--AI scientific work is designed becomes a central research question.

AI involvement in scientific inquiry takes different forms, with varying degrees of human direction and AI autonomy. In \emph{AI assistance}, human researchers retain scientific direction and consequential decisions while AI supports bounded activities such as literature retrieval, data analysis, or polishing scientific writing. Such assistance is already widespread: in a large survey of researchers at two German research organizations, about half reported using AI mainly to speed up their work, particularly on the tasks that take up most of their time \cite{ref08}. As AI capabilities develop, \emph{AI autonomy} allows systems to plan and execute longer portions of the research process with less continuous human intervention, which can extend the scale and speed of exploration \cite{ref01,ref02,ref05,ref06}. \emph{Human--AI Collaborative Science} (HACS) brings human judgment and AI capabilities into scientific inquiry through contributions that substantively shape its reasoning, decisions, and direction \cite{ref03,ref09,ref10}. This paper focuses on this collaborative mode.

HACS encompasses diverse arrangements of human and AI participation in scientific inquiry. Existing research provides important foundations for understanding these arrangements, including human and AI roles across scientific discovery \cite{ref09}, function allocation and delegation \cite{ref21,ref22,ref23}, team cognition and scientific criticism \cite{ref13,ref14,ref15,ref16,ref17,ref18,ref19,ref20}, and human--AI performance and interaction \cite{ref11,ref37}. These literatures offer complementary perspectives on how scientific work and authority are allocated, how participants coordinate, and how system performance is assessed. Yet the relationships among these aspects remain insufficiently articulated within an integrated account of human--AI scientific systems. Consequently, the organizational and coordination choices underlying existing HACS arrangements can be difficult to identify and compare.

Making these design relationships explicit raises three connected questions. First, when humans and AI participate in the same inquiry, who should do what, and who should determine what happens next? Second, how can humans and AI share and preserve the evolving state of scientific inquiry, carry contributions from proposal to resolution, and adapt their collaboration over time? Third, once such a collaborative system is in place, how should its effectiveness be evaluated, and how can performance differences be traced to its design? Together, these questions shift attention from the capabilities of individual humans or AI systems to the design and operation of the scientific system they jointly form.

\begin{figure}[!htb]
    \centering
    \includegraphics[width=0.78\linewidth]{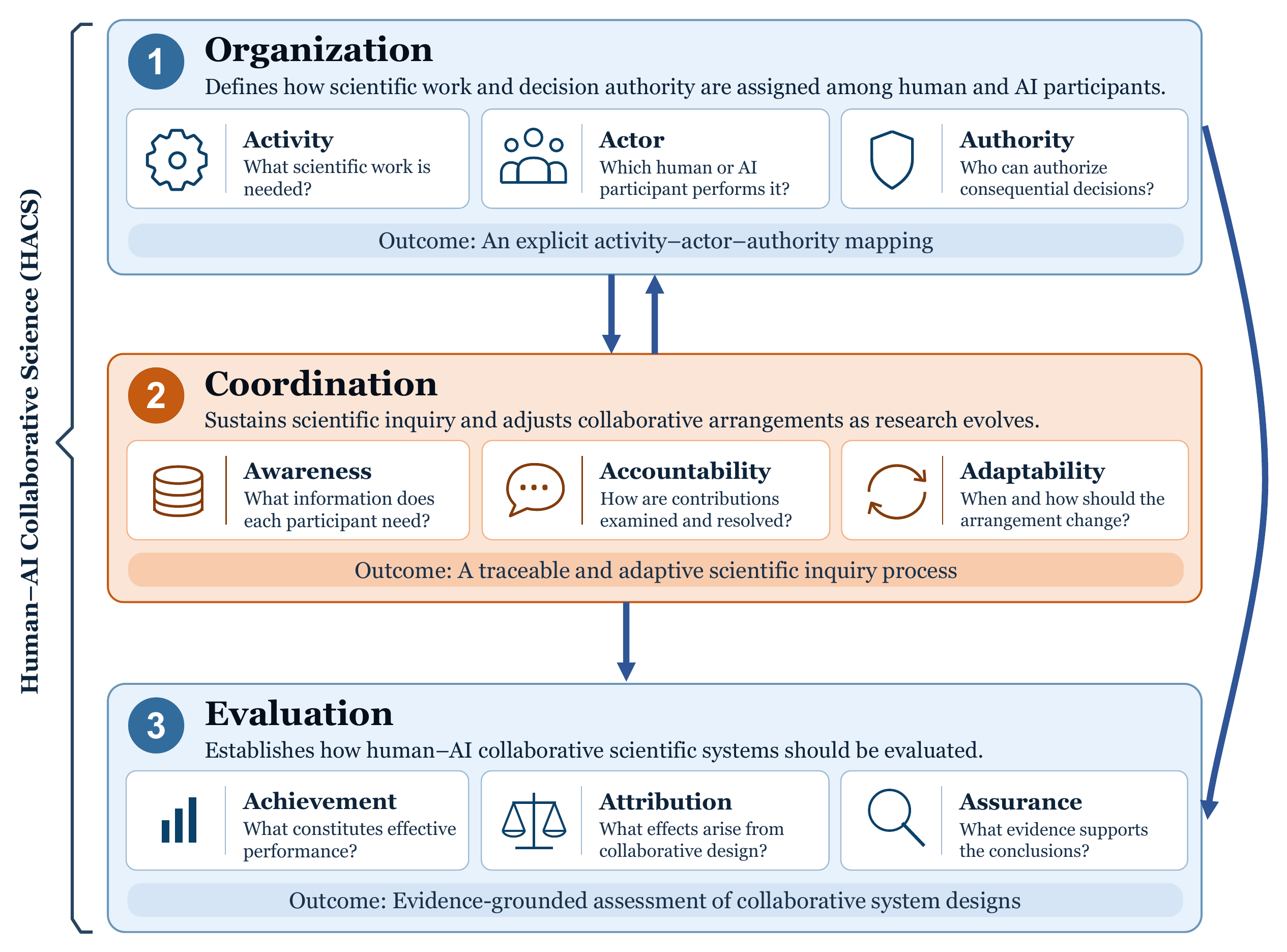}
    \caption{Three-layer framework for HACS. The organization layer specifies the activity--actor--authority mapping, the coordination layer supports the ongoing inquiry through shared information and review and adjusts collaborative arrangements as research evolves, and the evaluation layer examines achievement, attribution of design effects, and evidential assurance.}
    \label{fig:framework}
\end{figure}

We develop the \emph{9A Framework} for HACS to address these questions through three layers: Organization, Coordination, and Evaluation (Figure~\ref{fig:framework}). The organization layer specifies the scientific activities required (Activity), the human or AI participants who perform them (Actor), and the authority to make consequential decisions (Authority). The coordination layer addresses access to relevant research information (Awareness), the proposal, execution, examination, revision, and resolution of scientific contributions (Accountability), and adjustments to collaborative arrangements as research evolves (Adaptability). The evaluation layer defines criteria for assessing scientific system performance (Achievement), comparisons for identifying the effects of collaborative arrangements (Attribution), and the evidence required to support evaluation conclusions (Assurance). The three layers are analytically distinct but interconnected: organization defines collaborative arrangements, coordination sustains and adapts them, and evaluation assesses their performance and informs subsequent design choices.

The 9A Framework makes three contributions to the conceptualization and design of HACS. First, it defines HACS as a system-level design problem. Rather than focusing only on the capabilities of individual humans or AI agents, it identifies three essential aspects of collaborative scientific systems: Organization, Coordination, and Evaluation. Second, it connects previously separate research perspectives. Research on task allocation, human--AI coordination, and system evaluation often addresses different parts of the problem. The framework brings these perspectives together to explain how organizational choices shape ongoing collaboration and how their effects can be assessed. Third, it provides guidance for designing and improving collaborative scientific systems. Researchers can use the framework to specify how work and decision authority should be distributed, how scientific contributions should be coordinated and reviewed, and how alternative collaborative arrangements can be compared and refined.

The remainder of the paper reviews relevant foundations (Section~\ref{related-work-revised}), develops the three layers (Sections~\ref{organization-revised}--\ref{evaluation-revised}), and discusses their use, implications, and limitations (Section~\ref{discussion-revised}).

\section{Related Work}\label{related-work-revised}

Our framework draws on several literatures that address different parts of human--AI scientific work. After surveying AI-enabled scientific systems, we organize the remaining literatures around the three questions raised in the introduction: how scientific work and authority are allocated, how a collaborative scientific system remains coordinated over an evolving inquiry, and how the resulting system should be evaluated.

\subsection{AI-Enabled Scientific Systems and Collaborative Arrangements}
Representative AI-enabled scientific systems illustrate how human and AI participation is organized in practice, providing concrete cases for understanding different collaborative arrangements. Assistance tools support bounded activities under human direction, such as Elicit for literature search and evidence extraction and ChatGPT for language editing in scientific writing \cite{lagisz2026elicit,alyasiri2024chatgpt}. Systems with greater AI autonomy span multiple research stages: AI Scientist and AI Scientist-v2 connect idea generation, experimentation, analysis, and manuscript preparation; Robin connects AI-driven hypothesis generation and analysis with experimental biology; and autonomous chemistry agents and laboratories link computational reasoning with experimental planning and execution \cite{ref01,ref02,ref04,ref05,ref06}.

Collaborative systems vary in how human and AI contributions substantively shape the inquiry. Co-Scientist incorporates scientist-defined objectives and feedback into iterative hypothesis development \cite{ref03}; the Virtual Lab incorporates high-level human feedback on discussions among AI agents \cite{ref10}; and AutoResearchClaw offers intervention modes that differ in where and how often humans take part \cite{ref25}. Each system describes its arrangement in its own terms, such as specialized agents, agent roles and meetings, or intervention modes, which makes it difficult to compare how they distribute activities among actors and allocate authority.

\subsection{Organization and Delegation}
Human-factors and human--computer interaction research has long examined how functions should be allocated between people and automation. Parasuraman, Sheridan, and Wickens distinguish automation across information acquisition, information analysis, decision and action selection, and action implementation, with different levels of automation possible within each function \cite{ref21}. Shneiderman similarly argues that high levels of computer automation can coexist with high levels of human control \cite{ref22}. Work on human--agent teaming adds that joint activity depends on managing interdependence among participants \cite{ref26,ref27}. Together, these perspectives distinguish work allocation, degrees of automation, and human control.

Delegation research further examines the conditions under which tasks and decisions are assigned to AI. Lubars and Tan frame task delegability as a human-centered decision shaped by motivation, difficulty, risk, and trust \cite{ref23}. The Generative AI Delegation Taxonomy (GAIDeT) classifies AI contributions across stages of research and publishing while specifying degrees of human oversight \cite{suchikova2026gaidet}. More recent work considers the delegation of authority, responsibility, and accountability, together with explicit roles, boundaries, and intent \cite{ref28}. Te'eni, Nakash, and Drori propose a supervised workflow that embeds human oversight at every stage of AI-assisted research \cite{teeni2026oversight}. In scientific multi-agent systems, role labels such as planner, critic, analyst, or coder describe expected functions but may leave decision rights unspecified. A critic may identify a problem without having authority to stop the workflow. Recent work on collective and multi-agent AI calls for systematic attention to organizational structure, information flow, and role configuration \cite{ref29,ref30}. These literatures provide a foundation for examining work allocation and decision authority in collaborative scientific systems (Section~\ref{organization-revised}).

\subsection{Coordination in Scientific Inquiry}
When multiple participants contribute to the same scientific inquiry, coordination becomes a problem in its own right even when work allocation and decision boundaries are already specified. Coordination theory frames this problem as managing dependencies among activities and actors \cite{ref31}. Research on team cognition and computer-supported cooperative work has long examined how participants establish common ground, remain aware of one another's work, and sustain coordinated activity over time. Interactive Team Cognition, for example, treats team cognition as something that emerges through interaction at the team level \cite{ref13}. Related work on teamness, technology-as-teammate, grounding, and shared workspaces similarly emphasizes the importance of interaction, awareness, and external representations in collaborative work \cite{ref14,ref15,ref32,ref33}.

Coordination in scientific inquiry also involves the examination of competing ideas, evidence, and interpretations. Popper emphasizes criticism and severe testing as mechanisms for challenging scientific claims \cite{ref16}; Longino highlights the role of social interaction in exposing assumptions and values that may otherwise remain hidden \cite{ref17}; and Kitcher \cite{ref18} and Weisberg and Muldoon \cite{ref19} analyze the epistemic value of distributed search and division of cognitive labor. Related formal models examine how preserving diversity of beliefs or search strategies can improve collective inquiry under some conditions \cite{ref34,ref35,ref36}.

Recent human--AI research extends these concerns to feedback, expert judgment, and adaptive coordination in AI-enabled systems. Human--AI interaction guidelines and collaboration studies address correction, feedback, shared goals, responsiveness, and negotiated delegation \cite{ref37,ref38}. Studies of scientific judgment also examine how AI assessments align with expert evaluations. HybridQuestion finds stronger AI--expert agreement on established breakthroughs than on prospective research questions \cite{ref39}; \emph{AI Can Learn Scientific Taste} explores whether aspects of scientific significance can be learned from community feedback such as citation impact \cite{ref40}. Work on adjustable autonomy and adaptive systems further examines how decision-making control and system behavior can change with context \cite{ref41,ref42}. Together, these studies inform how collaborative scientific work is coordinated over time (Section~\ref{coordination}).

\subsection{Evaluation of Human--AI Scientific Systems}
Existing evaluations of AI-for-Science systems operate at several levels. End-to-end systems such as AI Scientist and AI Scientist-v2 are evaluated partly through the quality of their final research artifacts, including automated or human review of generated papers \cite{ref01,ref02}. Task-level benchmarks instead isolate specific scientific capabilities: ScienceAgentBench evaluates agents on 102 data-driven research tasks drawn from peer-reviewed studies, using generated programs, execution results, and cost as evaluation targets \cite{ref43}. More recent benchmarks move toward longer and more interactive settings: the SciAgentArena benchmark evaluates agents on real-world scientific scenarios with stepwise verification and an interactive environment \cite{ref44}.

These approaches have also exposed limitations in current evaluation practice. Artifact- or run-level success can conceal implementation drift, weak evidence use, and premature success declarations \cite{ref45,ref46}, and a score from a single selected implementation may not reliably represent the underlying idea \cite{ref47}. Task benchmarks such as ScienceAgentBench provide tighter control \cite{ref43}, but, as the SciAgentArena authors note, many reduce scientific work to static, well-specified problems with limited support for interactive evaluation \cite{ref44}. Work on multi-agent AI also calls for studying how organizational arrangements and design choices affect performance \cite{ref29,ref30}. Fan et al. describe factor-attribution comparisons that vary design choices while separating collaboration gains from gains due to added resources, providing a basis for investigating the effects of collaborative arrangements \cite{ref29}.

These approaches offer complementary ways to assess scientific outputs, task performance, and multi-agent design effects. Extending them to mixed human--AI scientific inquiry raises additional questions about human contributions and evidential quality (Section~\ref{evaluation-revised}).

\section{Organization: Activity, Actor, and Authority}\label{organization-revised}

Descriptions of AI-enabled scientific systems often identify participants through roles or architectural components \cite{ref09,ref48,ref49}. To understand their organization, we need to know which scientific activities are required, which participants perform them, and who can authorize consequential decisions. These questions are related, yet each captures a different organizational choice. The same activity may be assigned to a human researcher, an AI agent, or both; the actor performing it may also differ from the actor authorized to accept its results or redirect the inquiry. Even systems with similar participants can therefore have different organizational arrangements. We distinguish these choices through three concepts: \emph{activity}, \emph{actor}, and \emph{authority}. The following subsections develop each concept before examining how they relate.

\subsection{Activity}
Activity describes the scientific work that needs to be carried out in an inquiry. Depending on the problem, activities may include framing a question, retrieving evidence, generating hypotheses, designing a test, constructing a model, interpreting a result, searching for counterevidence, judging significance, or assessing whether a line of inquiry merits continuation. Empirical analyses of scientific workflows show variation in the activities used across scientific domains and applications \cite{garijo2014motifs}. The relevant activity set should therefore be specified for the inquiry under analysis.

Activity granularity should reflect the purpose of the analysis. Hierarchical task analysis treats the level of decomposition as a choice guided by what the analysis needs to reveal \cite{stanton2006hta}. A scientific modeling case study also documents trade-offs among activity decompositions at different granularities, including the added difficulty of finer-grained decomposition when iterative control is involved \cite{perrau2010granularity}. For our framework, the activity description should be detailed enough to distinguish consequential scientific judgments and assignments while remaining useful for comparing organizational arrangements.

\subsection{Actor}\label{sec:actor}
Actor identifies a participant---human, AI, or group---that performs activities or holds authority within an inquiry. An individual researcher or an AI agent may be represented as an actor, while a research team or a group of agents may also be treated as a single actor, depending on the purpose and granularity of the analysis. Such representation of groups as higher-level actors has precedents in holonic multi-agent systems \cite{gerber1999holonic}. When differences among group members are relevant to understanding activity allocation or decision authority, those members may be distinguished as separate actors. The appropriate level of representation therefore depends on what the analysis needs to make visible.

Actors may be characterized by roles such as principal investigator, critic, planner, or analyst, as commonly found in multi-agent systems \cite{zambonelli2000organisational}. In our framework, a role describes the general function of an actor rather than constituting a separate organizational element. For example, a critic agent may examine hypotheses, identify weaknesses in evidence, and suggest revisions. Such a role helps characterize the actor but does not fully specify its activities or decision authority, which are represented separately in the framework. Importantly, the relationship between activities and actors is not one-to-one: one actor may perform multiple activities, and multiple actors may contribute to the same activity.

\subsection{Authority}\label{sec:authority}
The third concept is \emph{authority}: who is permitted to turn a contribution or recommendation into a consequential decision or action that changes the scientific system. Authority is action-specific. An actor may be permitted to propose, recommend, execute, approve, redirect, stop, reassign, or escalate. The same concept applies whether the change concerns the inquiry itself or the organization of the collaboration; reassigning an activity or changing an authorization rule is simply another action over which authority must be assigned. Routine and reversible actions may be delegated broadly, while costly, risky, or difficult-to-reverse actions may require stronger authorization \cite{ref21,ref23,ref28}. The appropriate degree of delegation may also change as uncertainty and evidence quality change \cite{ref50}.

Formal authority can differ from effective influence \cite{ref51}. An actor that controls which hypotheses, evidence, or alternatives are presented can shape a decision without holding final approval rights \cite{ref52,ref53}. Likewise, human approval is substantive only when the authorizing person can inspect relevant evidence and objections, request further checks, and genuinely override or escalate a recommendation. Otherwise the checkpoint risks becoming ceremonial, particularly under automation bias \cite{ref54}, and the human may then be held responsible for outcomes they could not meaningfully control \cite{ref55,ref56}. Nor does supplying explanations or confidence scores by itself make oversight substantive: such information has not reliably improved joint performance \cite{ref11,ref57}. Analyzing an organization therefore requires attending to effective influence as well as formal authority.

\subsection{The Activity--Actor--Authority Mapping}
Activity, actor, and authority capture different aspects of scientific organization. To characterize a collaborative system, we must also specify how these concepts are connected. Research on levels of automation distinguishes the functions assigned to automation from the degree of human control over decisions and actions \cite{ref21}, while coactive design examines the interdependencies created when multiple participants contribute to joint work \cite{ref27}. These perspectives motivate examining the relationships among activities, participants, and decision rights together.

We represent these relationships through an \emph{activity--actor--authority mapping}. For each activity, the mapping identifies the actor or actors performing the work and the consequential decisions or actions associated with it, together with the actors authorized to make them. An actor may perform several activities, and several actors may contribute to the same activity. Authority can also differ across actions: generating a hypothesis, selecting it for further investigation, authorizing an experiment, and stopping that experiment need not be controlled by the same participant. This distinction is consistent with research on delegation that treats the scope of delegated decisions and responsibilities as an explicit design concern \cite{ref23,ref28}. Table~\ref{tab:organization-mapping-examples} illustrates how different activities and decisions within three scientific-agent systems involve different actors and authorizations.

\begin{table}[htbp]
\centering
\small
\setlength{\tabcolsep}{4pt}
\renewcommand{\arraystretch}{1.2}
\caption{Illustrative mappings in three scientific-agent systems. The activities, actors, and authorities are identified and characterized by the authors based on descriptions in the cited studies.}
\label{tab:organization-mapping-examples}
\begin{tabularx}{\linewidth}{@{}p{0.19\linewidth}p{0.22\linewidth}p{0.23\linewidth}X@{}}
\toprule
System & Activity & Actor & Authority \\
\midrule
Virtual Lab \cite{ref10} & Research direction and feedback & Human researcher & Sets the research objective and can redirect the inquiry through high-level feedback. \\
 & Agent discussions and scientific analysis & PI agent and scientist agents & PI agent directs agent discussions and coordinates subsequent agent work. \\
\addlinespace[3pt]
AutoResearchClaw \cite{ref25} & Research pipeline execution & AI agents & AI may advance stages autonomously; at designated checkpoints, a human researcher controls whether the workflow proceeds. \\
\addlinespace[3pt]
Co-Scientist \cite{ref03} & Hypothesis generation and ranking & Specialized AI agents & Human scientist selects which hypotheses proceed to experimental validation. \\
\bottomrule
\end{tabularx}
\end{table}

\section{Coordination: Awareness, Accountability, and Adaptability}\label{coordination}

Organization specifies how scientific work and decision rights are assigned at a given point, while inquiry unfolds through changing evidence, questions, and contributions. Sustaining collaboration across this process creates three connected needs. Participants must have access to the information relevant to their current responsibilities and decisions. Contributions must be proposed, examined, revised, and resolved in a way that preserves the evidence and reasoning behind what the inquiry retains. As research develops, persistent problems or changing requirements may also call for revisions to activity assignments or decision rights. We address these needs through \emph{awareness}, \emph{accountability}, and \emph{adaptability}, respectively. Together, they explain how an organizational arrangement supports ongoing scientific inquiry and how it can change over time.

\subsection{Awareness}\label{sec:awareness}
Awareness concerns how participants obtain the information relevant to their responsibilities as scientific inquiry evolves. Scientific work generates more information than any participant can usefully examine. A human researcher may need consequential findings, unresolved uncertainties, and decisions requiring attention, so that limited time and attention can be directed toward substantive scientific judgment rather than reviewing every intermediate result. Research on information overload documents the difficulties that arise when information demands exceed people's processing capacity \cite{eppler2004information}. A critic agent, by contrast, may need a claim, its supporting evidence, and the methods required to assess it, without necessarily receiving the original researcher's full interpretation. Such separation can leave room for alternative explanations and independent criticism; models of scientific communication show how restricted information exchange can sometimes preserve diversity in collective inquiry \cite{ref35,ref36}. Awareness therefore involves selecting information according to the scientific work each participant needs to perform.

Supporting these information needs across an evolving inquiry requires a persistent record of scientific progress and decisions. We use \emph{Research State} to describe a persistent, reconstructable representation of the inquiry, including research questions, hypotheses, evidence, methods, findings, uncertainties, objections, assignments, decisions, and provenance. Research State provides a consistent system-level information basis, while participants access the information relevant to their responsibilities rather than necessarily sharing an identical view. Related scientific-agent systems maintain operational state across research cycles \cite{ref49}; EviGraph represents scientific entities and their relationships in a typed evidence graph \cite{ref58}, while W3C PROV provides a foundation for describing provenance \cite{ref59}. In HACS, Research State also preserves organizational and decision context, including activity assignments and authorizations. Persistent records support continuity across research stages, while selective access allows participants to examine the same inquiry from different perspectives.

\subsection{Accountability}\label{sec:accountability}
Scientific inquiry develops through contributions that are proposed, tested, questioned, revised, and sometimes abandoned. Research on the social epistemology of science emphasizes critical interaction and responsiveness to criticism as conditions for collective inquiry \cite{ref17}. These contributions may introduce new hypotheses, evidence, methods, or objections that require further examination. For collaborative scientific work to progress, participants need to determine how such contributions are handled and preserve the reasoning behind consequential decisions. We use \emph{accountability} to describe this process: scientific contributions and the responses to them remain traceable, including what was examined, what concerns were raised, and why particular conclusions or actions were retained, revised, or rejected.

\begin{figure}[!htb]
    \centering
    \includegraphics[width=0.85\linewidth]{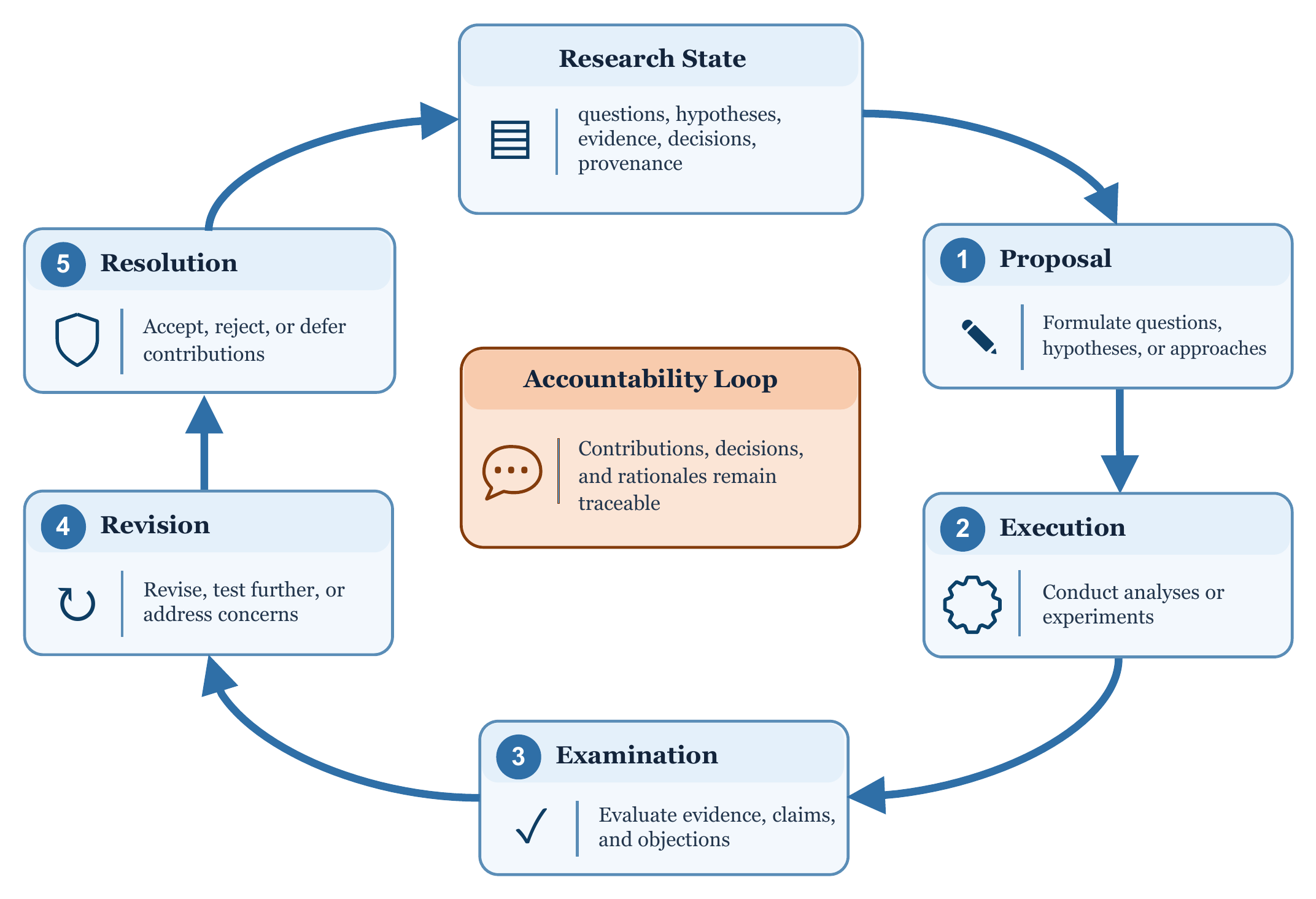}
    \caption{Dynamic coordination and scientific accountability. The diagram illustrates how scientific contributions move through proposal, execution, examination, revision, and resolution, supported by Research State.}
    \label{fig:coordination}
\end{figure}

We describe this process through a recurring \emph{Accountability Loop} (Figure~\ref{fig:coordination}). Participants propose approaches, execute analyses or experiments, examine the resulting evidence, and revise contributions when necessary. Each contribution can then be accepted, rejected, or deferred with a recorded rationale. Deferred contributions may remain unresolved until further evidence or analysis becomes available. These outcomes update Research State and inform subsequent inquiry. Consequential decisions follow the authority arrangements established in the organization layer, allowing scientific examination and decision authorization to be coordinated within the same process.

The Accountability Loop supports scientific progress by preserving how contributions and objections are examined and addressed. Participants can reconstruct what was attempted, which evidence supported a decision, what uncertainties remain, and why a particular research direction was pursued. Provenance models provide established ways to represent relationships among activities, agents, and resulting artifacts \cite{ref59}. The Virtual Lab's released meeting records illustrate one way collaborative scientific discussions can be documented and made available for subsequent examination \cite{ref60}. The depth of examination and documentation can vary with the significance and consequences of the scientific activity.

\subsection{Adaptability}\label{sec:adaptive-coordination}
Collaborative arrangements evolve as scientific inquiry progresses, participants gain experience, and new challenges emerge. Early in a research process, human researchers may closely review AI-generated results and approve consequential actions. As similar activities are repeated and their reliability is established through evaluation, some decisions may be delegated to AI agents with less frequent human intervention. Research on adjustable autonomy and adaptive systems provides a foundation for changing the degree of human involvement according to task conditions and system capabilities \cite{ref41,ref42}. At the same time, ongoing scientific work may reveal limitations in the existing arrangement. Repeated failures, unresolved methodological disagreements, or difficulties requiring specialized knowledge can indicate that current activity assignments or decision rights are no longer appropriate.

Adaptability concerns how these signals lead to revisions of the collaborative arrangement. Some difficulties can be addressed through ordinary scientific examination and revision, while persistent problems may require changes to the organization of subsequent work. For example, a participant may lack the tacit, local, or disciplinary knowledge needed for a task \cite{ref64}. Judging which open research questions merit pursuit is one example: AI assessments align less closely with expert judgment on prospective questions than on established breakthroughs \cite{ref39}, and current systems still struggle with long-horizon research \cite{ref45}. Such limitations may motivate bringing in human expertise, reassigning an activity, or changing approval requirements. Conversely, demonstrated improvements in AI capabilities, including the ability to learn aspects of scientific significance from community feedback \cite{ref40}, may support broader delegation when accompanied by appropriate validation. These adjustments can affect activities, actors, or decision authority, and they must follow applicable authorization requirements and be recorded in Research State so subsequent work reflects the revised arrangement.

\section{Evaluation: Achievement, Attribution, and Assurance}\label{evaluation-revised}

A human--AI scientific system may produce valuable research while consuming substantial computational resources or requiring extensive human review. Evaluating its practical value therefore begins with the scientific outcomes it achieves and the computational and human costs of achieving them. These measures describe system performance, which can also depend on the capabilities of the participants and the resources available. To understand the contribution of collaboration itself, we need to examine how choices in organization and coordination affect that performance. Finally, the strength of such conclusions depends on the evidence used to assess scientific quality and compare alternative arrangements, especially when research outcomes are open-ended or difficult to verify. These considerations motivate three evaluation concepts: \emph{achievement} measures performance, \emph{attribution} examines the effects of collaborative design, and \emph{assurance} assesses the credibility of the supporting evidence.

\subsection{Achievement}
Achievement concerns the value of a human--AI scientific system from the perspective of those using it to conduct research. We assess three dimensions: \emph{scientific outcome}, \emph{computational cost}, and \emph{human effort and experience}. Scientific outcome captures the quality and significance of the research produced. Computational cost captures the resources consumed by the AI system. Human effort and experience capture the time and attention required of researchers, alongside their experience of working with the system. These dimensions can vary independently: a system may produce strong scientific results while requiring substantial computation, extensive human repair, or disruptive oversight \cite{ref29,ref45}.

Table~\ref{tab:evaluation-dimensions} summarizes representative measures. The relevant measures depend on the scientific task: correctness may be directly testable in some settings, while novelty or significance may require expert judgment and later evidence. Resource reporting should include repeated attempts and hidden human correction where feasible. Human experience adds considerations such as cognitive workload, perceived control, and satisfaction, which elapsed time alone cannot capture \cite{hart1988nasatlx,ref37}. The three dimensions are assessed together without assuming a universal weighting among scientific value, machine resources, and human involvement.

\begin{table}[htbp]
\centering
\small
\setlength{\tabcolsep}{7pt}
\renewcommand{\arraystretch}{1.3}
\caption{Achievement dimensions for Human--AI Collaborative Science.}
\label{tab:evaluation-dimensions}
\begin{tabularx}{\linewidth}{@{}
  >{\raggedright\arraybackslash}p{0.23\linewidth}
  >{\raggedright\arraybackslash}p{0.29\linewidth}
  >{\raggedright\arraybackslash}X@{}}
\toprule
\textbf{Dimension} & \textbf{Core question} & \textbf{Example measures} \\
\midrule
Scientific outcome & What scientific value does the system produce? & Correctness where assessable, evidential support, robustness, novelty, significance \\
\addlinespace[3pt]
Computational cost & What machine resources does the system consume? & Tokens, model calls, runtime, compute usage, API costs \\
\addlinespace[3pt]
Human effort and experience & What does participation require of researchers, and how do they experience it? & Expert time, review workload, hidden repair, cognitive workload, perceived control, satisfaction \\
\bottomrule
\end{tabularx}
\end{table}

\subsection{Attribution}\label{sec:attribution}
We refer to the effects of organizational and coordination choices on the performance of human--AI scientific systems as the \emph{collaborative design effect}. This effect is assessed through the three achievement dimensions: scientific outcome, computational cost, and human effort and experience. Different arrangements can affect these dimensions in different ways. For example, assigning an independent critic to examine a research approach may reveal overlooked assumptions or methodological problems and improve scientific quality, while requiring additional computation. Providing participants with relevant information through Research State may reduce repeated analyses and unnecessary model calls. Concentrating human approval at consequential decision points may also reduce routine checking and allow researchers to focus on scientific judgment. Collaborative design effects can therefore involve improvements in research quality, reductions in resource demands, or better human experiences, alongside additional costs and trade-offs across dimensions.

Identifying collaborative design effects requires comparing system performance under alternative organizational and coordination arrangements. On the same or comparable scientific tasks, researchers can examine how different activity assignments, decision rights, or review mechanisms affect the three achievement dimensions. Such comparisons should account, as far as feasible, for differences in model capabilities, task difficulty, computational resources, and human contributions so that observed changes can be more plausibly associated with the collaborative design \cite{ref29,ref30}. For instance, to attribute a quality gain to an independent critic, a configuration with the critic can be compared against one without it that uses the same model and a comparable total computational budget, with the resources consumed by the critic itself counted toward that budget. This helps distinguish the effect of critique from improvements attributable to additional computation. Examining these effects and their trade-offs helps establish which arrangements are valuable under different research conditions and where the design of human--AI scientific systems can be improved.

\subsection{Assurance}\label{sec:validation}
Evaluations of human--AI scientific systems are not always straightforward to validate. Scientific inquiry often addresses open-ended questions whose outcomes cannot immediately be judged for correctness, novelty, or significance. Moreover, when system developers also evaluate their own systems, choices of metrics, successful cases, and interpretations may be influenced by expectations about the design. For example, reporting a large number of generated hypotheses does not by itself establish their scientific value. Similarly, a score obtained from one selected implementation does not by itself establish the merit of the underlying idea; assessing it requires comparison with independent implementations \cite{ref47}. These challenges make independent and credible evidence especially important. We use \emph{assurance} to describe the examination of how strongly the available evidence supports claims about achievement and the collaborative design effect.

Different scientific tasks call for different forms of evidence. For tasks with known outcomes, results can be checked against established answers, independently reproduced, or verified computationally. Open-ended tasks, such as generating hypotheses, discovering materials, or explaining poorly understood phenomena, may instead require independent expert assessment, experimental validation, and subsequent scientific evidence. The A-Lab case illustrates the importance of such scrutiny: the system initially reported the synthesis of 41 novel compounds from 58 targets \cite{ref06}, but subsequent examination challenged both the structural identification of the products and their novelty \cite{ref61}. A later correction clarified that the compounds were new to the prediction platform rather than necessarily new to science, and its manual reanalysis of the diffraction data confirmed 36 of the 40 syntheses retained in the corrected analysis, leaving four inconclusive \cite{ref62}. Reported scientific achievements therefore require evidence appropriate to the claims they make.

Assurance also concerns the scope of conclusions justified by that evidence. Strong performance on a small set of scientific tasks, for instance, does not establish that a particular collaborative arrangement will produce the same improvements across research domains. For the collaborative design effect, it is important to distinguish effects supported by the available evidence from uncertain observations and broader claims that have not been tested. Correctness, scientific significance, and the costs of collaboration may likewise be supported to different degrees. Making these distinctions explicit allows evaluations to communicate what has been established, under which conditions, and what remains to be validated.

\section{Discussion}\label{discussion-revised}

\subsection{Using the Framework}
The 9A Framework can support both the design of new human--AI scientific systems and the analysis of existing ones. When developing a workflow, researchers can make activity assignments, decision rights, information sharing, and review procedures explicit, and expose selected organizational and coordination choices as configurable parameters for different scientific tasks. For an existing system, the framework provides a way to diagnose specific collaboration problems. Repeated literature searches may point to inadequate information sharing; excessive human approvals may indicate poorly allocated decision rights; and scientific objections that remain unresolved may reveal gaps in accountability. Identifying the organizational or coordination choices behind such problems can inform targeted changes, whose consequences can then be examined through achievement and the collaborative design effect.

An open-source research workflow (\url{https://github.com/ZihanZheng2000/human-ai-collaborative-science-framework}) illustrates one implementation of these ideas, including explicit decision rights, Research State, and review loops. It provides a concrete example of how the framework's organizational and coordination concepts can be instantiated in a scientific workflow.

\subsection{Broader Implications}
Over the longer term, human--AI collaboration may change how scientific research is organized. As AI systems take on a wider range of research activities, human researchers may devote more attention to selecting research directions, judging scientific evidence, and making consequential decisions. The division of activities, decision authority, and collaborative relationships may also evolve with system capabilities, research needs, and stages of inquiry. At the same time, these collaborative arrangements are shaped by institutional requirements, including research ethics, funding conditions, publication policies \cite{ref68,ref69,ref70}, and norms of scientific responsibility.

These changes in scientific organization also have implications for responsibility and transparency. Existing publishing guidance places scholarly accountability primarily with human authors, reviewers, and editors \cite{ref68,ref69,ref70}, while contributor-role descriptions such as CRediT may not capture the authorization of consequential research decisions \cite{ref71}. Hosseini et al. further argue that research evaluation should recognize human contributions involving situated judgment, accountability, and sustained scholarly relationships \cite{hosseini2026evaluation}. Important findings may emerge through successive proposals, challenges, revisions, and approvals that are only partially visible in the final publication. This raises questions about how scientific contributions and decisions should be documented, attributed, and evaluated when both humans and AI participate in research.

\subsection{Scope and Limitations}
The framework is conceptual, and its proposed relationships have yet to be tested in real scientific settings. Scientific domains differ in research duration, task risks, available evidence, and dependence on expert judgment, so appropriate activity granularity, authorization thresholds, and coordination intensity may vary across contexts. Explicit information sharing, review, and record-keeping also impose computational and human costs. For bounded, low-risk, and easily verifiable tasks, simple AI assistance or predominantly autonomous execution may be more appropriate than elaborate collaborative arrangements. More broadly, widespread reliance on similar AI models across research teams may also raise concerns about convergence in scientific questions, methods, and research directions. Such collective effects raise broader questions of epistemic responsibility \cite{ref20}.

\section{Conclusion}\label{conclusion-revised}

This paper conceptualizes HACS as a scientific system whose effectiveness depends on how human and AI participants organize and coordinate their contributions. The 9A Framework provides an integrated perspective through three layers: organization, coordination, and evaluation.

By connecting the allocation of scientific activities and decision authority with the coordination of ongoing inquiry, the framework offers a basis for designing and analyzing collaborative scientific systems. Its evaluation layer further enables the examination of scientific outcomes, computational costs, and human effort and experience, as well as the effects of collaborative design and the evidence supporting these assessments.

As AI takes on a growing range of scientific activities, the organization of human--AI research will become an increasingly important design consideration. The framework provides a foundation for examining which collaborative arrangements are effective, under what conditions, and at what cost. Establishing these relationships through empirical research is an important next step toward more effective and accountable human--AI scientific inquiry.

\bibliographystyle{unsrturl}
\bibliography{references}

\end{document}